\documentclass[aps,prd,twocolumn,nofootinbib,showpacs]{revtex4}

\usepackage{graphicx}

\usepackage[colorlinks=true, linkcolor=blue, citecolor=blue, urlcolor=blue]{hyperref}

\begin{document}

\title{Novel method for the precise determination of the QCD running coupling from event shape distributions in electron-positron annihilation}

\author{Sheng-Quan Wang$^{1,2}$}
\email[email:]{sqwang@cqu.edu.cn}

\author{Stanley J. Brodsky$^2$}
\email[email:]{sjbth@slac.stanford.edu}

\author{Xing-Gang Wu$^3$}
\email[email:]{wuxg@cqu.edu.cn}

\author{Jian-Ming Shen$^{4}$}
\email[email:]{cqusjm@cqu.edu.cn}

\author{Leonardo Di Giustino$^{2,5}$}
\email[email:]{leonardo.digiustino@gmail.com}

\address{$^1$Department of Physics, Guizhou Minzu University, Guiyang 550025, P.R. China}
\address{$^2$SLAC National Accelerator Laboratory, Stanford University, Stanford, California 94039, USA}
\address{$^3$Department of Physics, Chongqing University, Chongqing 401331, P.R. China}
\address{$^4$School of Physics and Electronics, Hunan University, Changsha 410082, P.R. China}
\address{$^5$Department of Science and High Technology, University of Insubria, via valleggio 11, I-22100, Como, Italy}

\begin{abstract}

We present a novel method for precisely determining the running QCD coupling constant $\alpha_s(Q^2)$ over a wide range of $Q^2$ from event shapes for electron-positron annihilation measured at a single annihilation energy $\sqrt{s}$.  The renormalization scale $Q^2$ of the running coupling depends dynamically on the virtuality of the underlying quark and gluon subprocess and thus the specific kinematics of each event.  The determination of the renormalization scale for event shape distributions is obtained by using the Principle of Maximum Conformality (PMC), a rigorous scale-setting method for gauge theories which satisfies all the requirements of Renormalization Group Invariance, including renormalization-scheme independence and consistency with Abelian theory in the $N_C \to 0$ limit.  In this paper we apply the PMC to two classic event shapes measured in $e^+ e^-$ annihilation:  the thrust ($T$) and C-parameter ($C$).  The PMC renormalization scale depends differentially on the values of $T$ and $C$.  The application of PMC scale-setting determines the running coupling $\alpha_s(Q^2)$ to high precision over a wide range of $Q^2$ from $10$ to $250~\rm{GeV}^2$ from measurements of the event shape distributions at the $Z^0$ peak.  The extrapolation of the running coupling using pQCD evolution gives the value $\alpha_s(M^2_Z)=0.1185\pm0.0012$ from the thrust, and $\alpha_s(M^2_Z)=0.1193^{+0.0021}_{-0.0019}$ from the C-parameter in the $\overline {\rm{MS}}$ scheme. These determinations of $\alpha_s(M^2_Z)$ are consistent with the world average and are more precise than the values obtained from analyses of event shapes currently used in the world average.  The highly-consistent results for the $T$ and $C$ event-shape distributions provide an additional verification of the applicability of the PMC to pQCD.

\end{abstract}

\maketitle

\section{Introduction}

The strong coupling constant, $\alpha_s(Q^2)$, is the fundamental coupling underlying Quantum Chromodynamics (QCD) and its predictions for hadron and nuclear physics.  It is thus crucial to determine $\alpha_s(Q^2)$ to the best possible precision.  The dependence of  $\alpha_s(Q^2)$  on the renormalization scale $Q^2$ obtained from many different physical processes show consistency with QCD predictions and asymptotic freedom. The Particle Data Group (PDG) currently gives the world average: $\alpha_s(M^2_Z)=0.1181\pm0.0011$~\cite{Tanabashi:2018oca} in the $\overline{\rm{MS}}$ renormalization scheme.

An important test of the consistency of the QCD predictions can be obtained from the analysis of event shapes in electron-positron annihilation.
A precise determination of $\alpha_s(Q^2)$ can be obtained  from a detailed comparison of the theoretical predictions with the experimental data, especially by using the large data sample available at the $Z^0$ peak. In fact, the main obstacle for achieving a highly precise determination of the QCD coupling from event shapes is not the lack of precise experimental data, but the ambiguity of theoretical predictions.

Currently, theoretical calculations for event shapes are based on ``conventional" scale setting; i.e., one simply sets the value of the renormalization scale equal to the center-of-mass energy $\mu_r=\sqrt{s}$; the theory uncertainties for this guess are estimated by varying the renormalization scale over an arbitrary range; e.g., $\mu_r\in[\sqrt{s}/2,2\sqrt{s}]$. By using conventional scale setting, only one value of $\alpha_s$ at the scale $\sqrt{s}$ can be extracted, and the main source of the uncertainty is the choice of the renormalization scale. For example, the value of $\alpha_s(M^2_Z)=0.1224\pm 0.0039$~\cite{Dissertori:2009ik}, with a perturbative uncertainty of $0.0035$, is obtained by using next-to-next-to-leading order (NNLO)+next-to-leading-logarithmic approximation (NLLA) predictions. Recent determinations of $\alpha_s$ based on the soft-collinear effective theory are $\alpha_s(M^2_Z)=0.1135\pm0.0011$~\cite{Abbate:2010xh} from the thrust, and $\alpha_s(M^2_Z)=0.1123\pm0.0015$~\cite{Hoang:2015hka} from the C-parameter.  Theorists have introduced corrections, such as non-perturbative hadronization effects, in order to match the theoretical predictions to the experimental data.  However, as pointed out in Ref.\cite{Tanabashi:2018oca}, the systematics of the theoretical uncertainties for extracting $\alpha_s$ using Monte Carlo generators to simulate the non-perturbative hadronization effects are not well understood.

Conventional scale setting introduces an inherent scheme-and-scale dependence for pQCD predictions, and it violates a fundamental principle of Renormalization Group Invariance (RGI): theoretical predictions cannot depend on arbitrary conventions such as the renormalization scheme.   One often argues that the inclusion of higher-order terms will suppress the scale uncertainty; however, estimating unknown higher-order terms by simply varying the renormalization scale within an arbitrary range is unreliable since it is only sensitive to the $\beta$ terms.  In fact, the resulting pQCD series diverges strongly as $\alpha_s^n \beta_0^n n!$, the ``renormalon" divergence~\cite{Ellis:1995jv}.  Moreover, the conventional procedure of guessing the renormalization scale is inconsistent with the Gell-Mann-Low procedure~\cite{GellMann:1954fq} which determines the scale unambiguously in QED.  pQCD predictions must analytically match Abelian theory  in the $N_C\rightarrow 0$ limit~\cite{Brodsky:1997jk}.

The Principle of Maximum Conformality (PMC)~\cite{Brodsky:2011ta, Brodsky:2012rj, Brodsky:2011ig, Mojaza:2012mf, Brodsky:2013vpa} provides a systematic way to eliminate the renormalization scheme-and-scale ambiguities. The PMC scales are fixed by absorbing the $\beta$ terms that govern the behavior of the running coupling via the Renormalization Group Equation (RGE).  Since the PMC predictions do not depend on the choice of the renormalization scheme, PMC scale setting satisfies the principles of RGI~\cite{Brodsky:2012ms, Wu:2014iba, Wu:2019mky}.  Since the $\beta$ terms do not appear in the pQCD series after the PMC, there is no renormalon divergence. The PMC method extends the Brodsky-Lepage-Mackenzie (BLM) scale-setting method~\cite{Brodsky:1982gc} to all orders, and it reduces in the Abelian limit to the Gell-Mann-Low method~\cite{GellMann:1954fq}.

In this paper, we will apply the PMC to make comprehensive analyses for two classic event shapes: the thrust ($T$)~\cite{Brandt:1964sa, Farhi:1977sg} and the C-parameter ($C$)~\cite{Parisi:1978eg, Donoghue:1979vi}. The PMC renormalization scale depends dynamically on the virtuality of the underlying quark and gluon subprocess and thus the specific kinematics of each event.  We then can determine $\alpha_s(Q^2)$ over a large range of $Q^2$ by comparing the PMC predictions with the experimental data.

\section{Numerical results and discussions for the thrust and C-parameter}

The thrust and C-parameter are defined as
\begin{eqnarray}
T=\max\limits_{\vec{n}}\left(\frac{\sum_{i}|\vec{p}_i\cdot\vec{n}|}{\sum_{i}|\vec{p}_i|}\right),
\end{eqnarray}
\begin{eqnarray}
C=\frac{3}{2}\frac{\sum_{i,j}|\vec{p_i}||\vec{p_j}|\sin^2\theta_{ij}}{\left(\sum_i|\vec{p_i}|\right)^2},
\end{eqnarray}
where $\vec{p}_i$ denotes the three-momentum of particle $i$. For the thrust, the unit vector $\vec{n}$ is varied to define the thrust direction $\vec{n}_T$ by maximizing the sum on the right-hand side. For the C-parameter, $\theta_{ij}$ is the angle between $\vec{p_i}$ and $\vec{p_j}$. The range of values is $1/2\leq T\leq1$ for the thrust, and for the C-parameter it is $0\leq C\leq1$.

For our numerical computations, we use the EVENT2 program~\cite{Catani:1996jh} to precisely calculate the perturbative coefficients at the next-to-leading order (NLO). The perturbative coefficients at the next-to-next-to-leading order (NNLO) can be calculated using the EERAD3 program~\cite{Ridder:2014wza}, and are checked using the results of Ref.\cite{Weinzierl:2009ms}. We use the RunDec program~\cite{Chetyrkin:2000yt} to evaluate the $\overline{\rm MS}$ scheme running coupling from $\alpha_s(M_Z)=0.1181$~\cite{Tanabashi:2018oca}.

\begin{figure}[htb]
\centering
\includegraphics[width=0.35\textwidth]{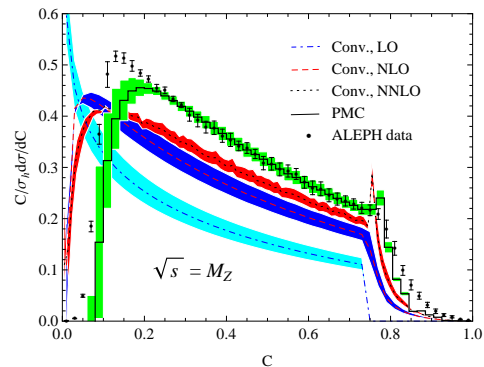}
\caption{The C-parameter differential distributions using conventional (Conv.) and PMC scale settings at $\sqrt{s}=M_Z$. The dot-dashed, dashed and dotted lines are the conventional scale-fixed results at LO, NLO and NNLO~\cite{Ridder:2014wza, Weinzierl:2009ms}, respectively, and the corresponding error bands are obtained by varying $\mu_r\in[M_Z/2,2M_Z]$. The solid line is the PMC result, and its error band is the squared averages of the errors for $\alpha_s(M_Z)=0.1181\pm0.0011$~\cite{Tanabashi:2018oca} and the estimated unknown higher-order contributions $\pm 0.2$ $C_n$. The data are taken from the ALEPH~\cite{Heister:2003aj} experiment.}
\label{figuseConvC}
\end{figure}

A detailed PMC analysis for the thrust has been given in Ref.\cite{Wang:2019ljl}. We calculate the C-parameter following a similar procedure and present its differential distributions at $\sqrt{s}=M_Z$ in Fig.(\ref{figuseConvC}). Figure(\ref{figuseConvC}) shows that the conventional predictions -- even up to NNLO pQCD corrections -- substantially deviate from the precise experimental data.  The conventional predictions are plagued by the scale uncertainty.  Since the variation of the scale is only sensitive to the $\beta$ terms, the estimate of unknown higher-order terms by varying $\mu_r\in[\sqrt{s}/2,2\sqrt{s}]$ is unreliable:  the NLO calculation does not overlap with the LO prediction, and the NNLO calculation does not overlap with NLO prediction.   In addition, the perturbative series for the C-parameter distribution shows slow convergence because of the renormalon divergence. In contrast, Fig.(\ref{figuseConvC}) shows that PMC prediction for the C-parameter distribution is in excellent agreement with the experimental data.  There is some deviation near the two-jet and multi-jet regions, which is expected since pQCD becomes unreliable due to the presence of large logarithms in those kinematic regions. The resummation of large logarithms is thus required, and this topic has been extensively studied in the literature.

It should be emphasized that the PMC eliminates the scale $\mu_r$ uncertainty; the conventional estimate of unknown higher-order terms obtained by varying $\mu_r\in[\sqrt{s}/2,2\sqrt{s}]$ is not applicable to the PMC predictions.  An estimate of the unknown higher-order contributions can be characterized by the convergence of the perturbative series and the magnitude of the last-known higher-order term.  We note that the relative magnitude of the corrections for the C-parameter distribution is $C_{\rm LO}:C_{\rm NLO}:C_{\rm NNLO}\sim1:0.5:0.2$~\cite{GehrmannDeRidder:2007hr} in the intermediate region using conventional scale setting.  After using the PMC, the relative magnitude at NLO is improved to be $C_{\rm LO}:C_{\rm NLO}\sim1:0.2$. The error estimate of an $n$th-order calculation can be characterized by the last known term;  i.e., $\pm C_n$, where $n$ stands for LO, NLO, NNLO, $\cdots$.  After applying the PMC, the unknown $C_{n+1}$ term can be estimated using $\pm ~0.2~C_n$ if one assumes that the relative magnitude of the unknown $(n+1)$th-order term is the same as that of the known $n$th-order term; i.e., $C_{n+1}/C_{n}=C_{n}/C_{n-1}$. The resulting PMC error bar for the C-parameter distribution is presented in Fig.(\ref{figuseConvC}). This estimate of the unknown higher-order terms is natural for a convergent perturbative series.

\begin{figure}[htb]
\centering
\includegraphics[width=0.35\textwidth]{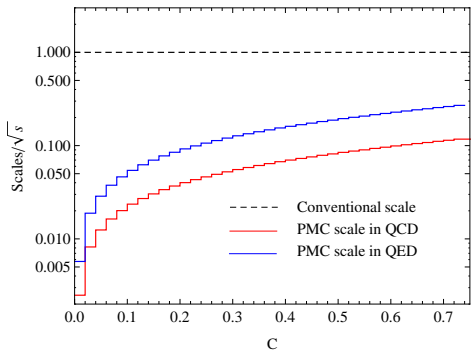}
\caption{The PMC scale for the C-parameter. As a comparison, the scale $\mu_r=\sqrt{s}$ using conventional scale-setting and the PMC scale in QED are also presented. }
\label{figPMCscale}
\end{figure}

Unlike conventional scale-setting, where the scale is fixed at $\mu_r=\sqrt{s}$, the PMC scale is determined by absorbing the $\beta$ terms of the pQCD series into the coupling constant. The resulting PMC scale is not a single value, but it monotonously increases with the value of $C$, reflecting the increasing virtuality of the QCD dynamics.  Thus, simply fixing the scale at $\mu_r=\sqrt{s}$ obviously violates the physical behavior of the C-parameter distribution.   In addition, the number of active flavors $n_f$ changes with the value of $C$ according to the PMC scale.  More explicitly, the PMC scale in the $0<C<0.75$ region is presented in Fig.(\ref{figPMCscale}). The LO contribution vanishes in the $0.75<C<1$ region; the NLO PMC scale is determined in this domain by using the NNLO contribution.  Near the two-jet region, the quarks and gluons have soft virtuality, and the PMC renormalization scale becomes small. The pQCD theory thus becomes unreliable in this domain. The dynamics of the PMC scale thus signals the correct physical behavior in the two-jet region.  After PMC scale-setting, the resulting pQCD series with $\beta=0$ gives the prediction for a ``conformal collider"~\cite{Hofman:2008ar}.  The correct physical behavior of the scale for event shapes was also obtained in Refs.\cite{Kramer:1990zt, Gehrmann:2014uva}. Soft-collinear effective theory also determines the C-parameter distribution at different energy scales~\cite{Hoang:2014wka}.

\begin{figure}[htb]
\centering
\includegraphics[width=0.35\textwidth]{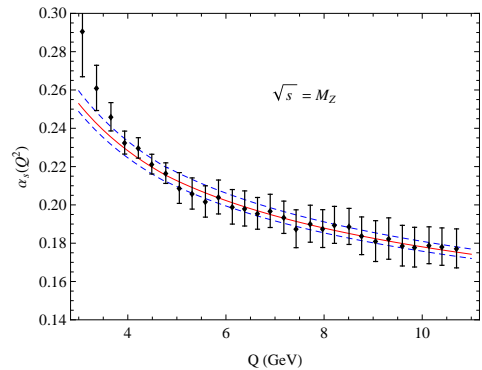}
\caption{The coupling constant $\alpha_s(Q^2)$ extracted by comparing PMC predictions with the ALEPH data~\cite{Heister:2003aj} at a single energy of $\sqrt{s}=M_Z$ from the C-parameter distributions in the $\overline{\rm MS}$ scheme. The error bars are the squared averages of the experimental and theoretical errors. The three lines are the world average evaluated from $\alpha_s(M^2_Z)=0.1181\pm 0.0011$~\cite{Tanabashi:2018oca}. }
\label{figasPMCT}
\end{figure}

Since the renormalization scale is simply set as $\mu_r=\sqrt{s}$ when using conventional scale setting, only one value of $\alpha_s$ at scale $\sqrt{s}$ can be extracted. In contrast, since the PMC scale varies with the value of the event shape $C$, we can extract $\alpha_s(Q^2)$ over a wide range of $Q^2$ using the experimental data at a single energy of $\sqrt{s}$. By adopting a method similar to~\cite{Becher:2008cf}, we have determined $\alpha_s(Q^2) $ bin-by-bin from the comparison of PMC predictions with measurements at $\sqrt{s}=M_Z$; see Fig.(\ref{figasPMCT}). The results for $\alpha_s(Q^2)$ in the range $3$ GeV $<Q<11$ GeV are in excellent agreement with the world average evaluated from $\alpha_s(M^2_Z)$~\cite{Tanabashi:2018oca}.  Since the PMC method eliminates the renormalization scale uncertainty, the extracted $\alpha_s(Q^2)$ is not plagued by any uncertainty from the choice of $\mu_r$. The results for $\alpha_s(Q^2)$ obtained from the thrust observable using the PMC are consistent with the results using the C-parameter~\cite{Wang:2019ljl}. Thus, PMC scale-setting provides a remarkable way to verify the running of $\alpha_s(Q^2)$ from event shapes measured at a single energy of $\sqrt{s}$.

\section{Mean values for the thrust and C-parameter}

The differential distributions of event shapes are afflicted with large logarithms in the two-jet region. The comparison of QCD predictions with experimental data and then extracting $\alpha_s$ are restricted to the region where leading-twist pQCD theory is able to describe the data well. Choosing different domains of the distributions leads to different values of $\alpha_s$. Note that the mean value of event shapes,
\begin{eqnarray}
\langle y\rangle &=&\int_0^{y_0}\frac{y}{\sigma_{h}}\frac{d\sigma}{dy}dy,
\end{eqnarray}
where $y_0$ is the kinematically allowed upper limit of the $y$ variable, involves an integration over the full phase space, it thus provides an important complement to the differential distributions and to determinate $\alpha_s$.

The PMC renormalization scales corresponding to the mean values for the thrust and C-parameter are
\begin{eqnarray}
\mu^{\rm{pmc}}_r|_{\langle1-T\rangle} = 0.0695\sqrt{s}, ~\rm{and}~ \mu^{\rm{pmc}}_r|_{\langle C\rangle} = 0.0656 \sqrt{s}, \nonumber
\end{eqnarray}
respectively. The PMC scales satisfy $\mu^{\rm pmc}_r\ll \sqrt{s}$ reflecting the virtuality of the underlying QCD subprocesses and the effective number of quark flavors $n_f$.
We note that the analysis of Ref.\cite{Heister:2003aj} using conventional scale setting leads to an anomalously large value of $\alpha_s$, demonstrating again that the correct description for the mean values requires $\mu_r\ll \sqrt{s}$.

In the case of the center-of-mass energy at the $Z^0$ peak, $\sqrt{s}=M_Z=91.1876$ GeV, the PMC scales are $\mu^{\rm pmc}_r|_{\langle1-T\rangle}=6.3$ GeV and $\mu^{\rm pmc}_r|_{\langle C\rangle}=6.0$ GeV for the thrust and C-parameter, respectively. The PMC scales of the differential distributions for the thrust and C-parameter are also very small. The average of the PMC scales $\langle\mu^{\rm pmc}_r\rangle$ of the differential distributions for the thrust and C-parameter are close to the PMC scales $\mu^{\rm pmc}_r|_{\langle1-T\rangle}$ and $\mu^{\rm pmc}_r|_{\langle C\rangle}$, respectively. This shows that PMC scale setting is self-consistent with the differential distributions for the event shapes and their mean values.

\begin{figure}[htb]
\centering
\includegraphics[width=0.35\textwidth]{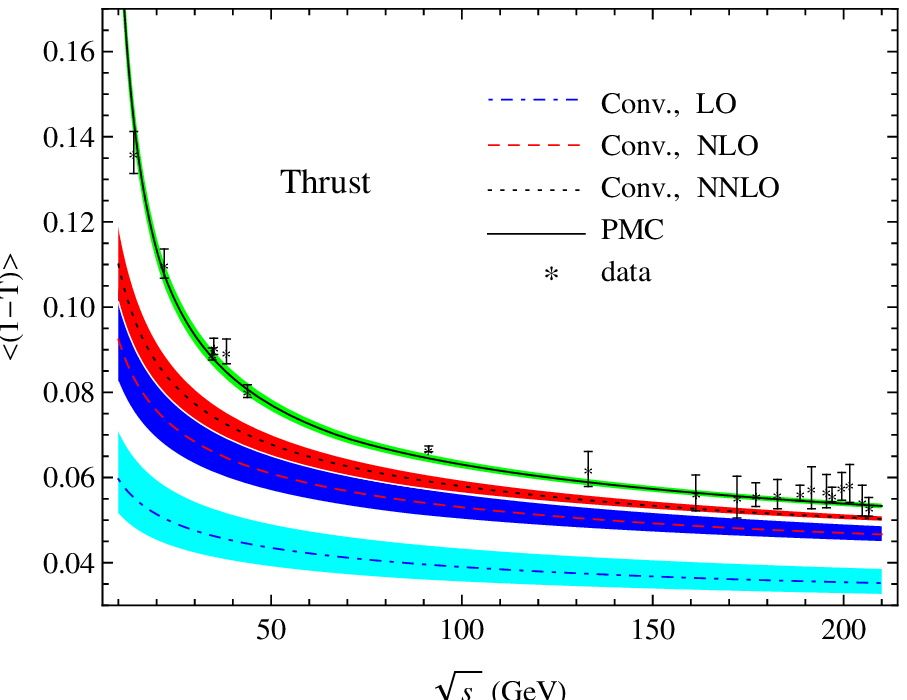}
\includegraphics[width=0.35\textwidth]{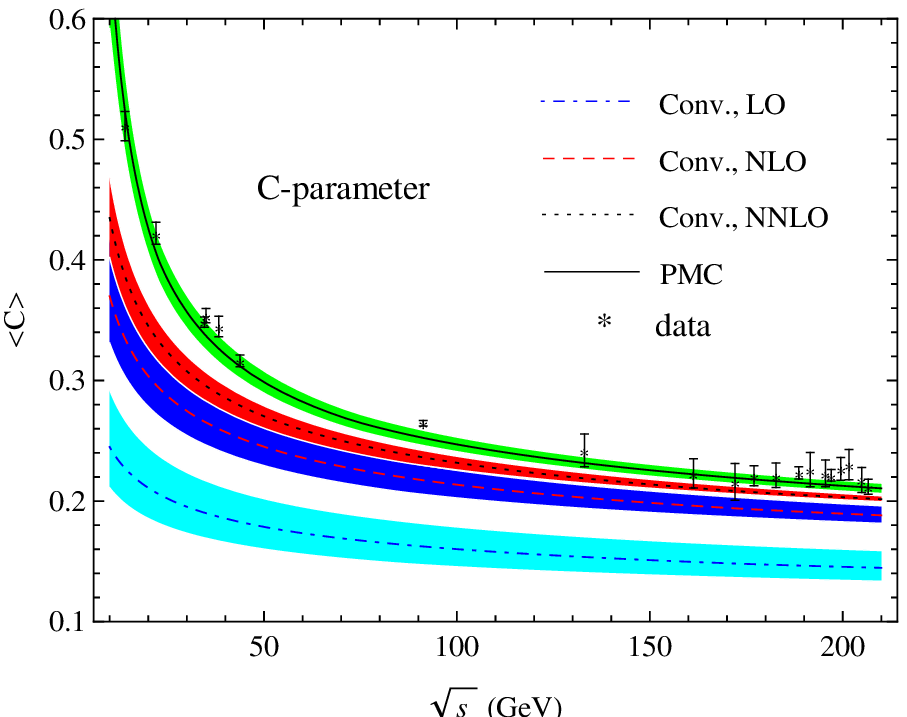}
\caption{The mean values for the thrust (up) and C-parameter (down) versus the center-of-mass energy $\sqrt{s}$ using conventional (Conv.) and PMC scale settings. The dot-dashed, dashed and dotted lines are the conventional results at LO, NLO and NNLO~\cite{GehrmannDeRidder:2009dp, Weinzierl:2009yz}, respectively, and the corresponding error bands are obtained by varying $\mu_r\in[M_Z/2,2M_Z]$. The solid line is the PMC result, and its error band is obtained by the squared averages of the errors for $\alpha_s(M_Z)=0.1181\pm0.0011$~\cite{Tanabashi:2018oca} and the estimated unknown higher-order contributions $\pm 0.2$ $C_n$. The data are from the JADE and OPAL experiments, taken from~\cite{Abbiendi:2004qz, Pahl:2007zz}. }
\label{convPMCmoment}
\end{figure}

We present the mean values for the thrust and C-parameter versus the center-of-mass energy $\sqrt{s}$ in Fig.(\ref{convPMCmoment}). It shows that in the case of conventional scale setting, the predictions are plagued by the renormalization scale $\mu_r$ uncertainty and substantially deviate from measurements even up to NNLO~\cite{GehrmannDeRidder:2009dp, Weinzierl:2009yz}.  In contrast, after using PMC scale setting, the mean values for the thrust and C-parameter are increased, especially for  small $\sqrt{s}$.  The scale-independent PMC predictions are in excellent agreement with the experimental data over the wide range of center-of-mass energies $\sqrt{s}$. Thus, PMC scale setting provides a rigorous, comprehensive description of the measurements without artificial parameters.

\begin{figure}[htb]
\centering
\includegraphics[width=0.35\textwidth]{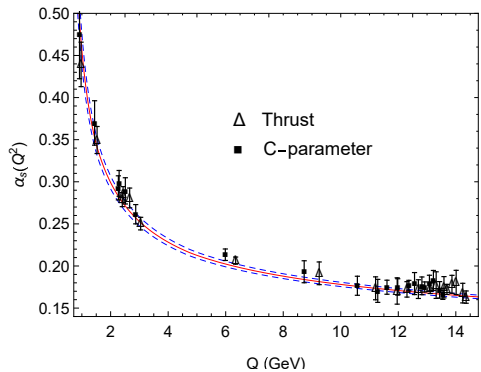}
\caption{The running coupling $\alpha_s(Q^2)$ extracted from the thrust and C-parameter mean values by comparing PMC predictions with the JADE and OPAL data~\cite{Abbiendi:2004qz, Pahl:2007zz} in the $\overline{\rm MS}$ scheme. The error bars are the squared averages of the experimental and theoretical errors. The three lines are the world average evaluated from $\alpha_s(M^2_Z)=0.1181\pm0.0011$~\cite{Tanabashi:2018oca}. }
\label{convPMCmomentas}
\end{figure}

Since a high degree of consistency between the PMC predictions and the measurements is obtained, we can extract $\alpha_s(Q^2)$ with high precision; the results in the $\overline{\rm MS}$ scheme are presented in Fig.(\ref{convPMCmomentas}). The values obtained for $\alpha_s(Q^2)$ are mutually compatible and are in excellent agreement with the world average in the range $1$ GeV $<Q<15$ GeV. The results are not plagued by the renormalization scale $\mu_r$ uncertainty. In addition, unlike the $\alpha_s$ extracted from the differential distributions, the $\alpha_s$ extracted from the mean values are not afflicted with large logarithmic contributions nor non-perturbative effects.

We can also obtain a highly precise determination of the value of $\alpha_s(M^2_Z)$ from a fit of the PMC predictions to the measurements. We adopt the method similar to~\cite{Pahl:2008uc} and the $\chi^2$-fit is defined by $\chi^2 = \sum_{i}\left((\langle y\rangle^{\rm {exp.}}_i - \langle y\rangle^{\rm {theo.}}_i)/\sigma_i \right)^2$, where $\langle y\rangle^{\rm {exp.}}_i$ is the value of the experimental data, $\sigma_i$ is the corresponding experimental uncertainty, $\langle y\rangle^{\rm {theo.}}_i$ is the theoretical prediction. The $\chi^2$ value is minimized with respect to $\alpha_s(M^2_Z)$ for the thrust and C-parameter separately.  We obtain
\begin{eqnarray}
\alpha_s(M^2_Z) &=& 0.1185\pm0.0011(\rm Exp.)\pm0.0005(\rm Theo.) \nonumber \\
&& = 0.1185\pm0.0012,
\end{eqnarray}
with $\chi^2/$d.o.f.$=27.3/20$ for the thrust mean value, and
\begin{eqnarray}
\alpha_s(M^2_Z) &=& 0.1193^{+0.0009}_{-0.0010}(\rm Exp.)^{+0.0019}_{-0.0016}(\rm Theo.) \nonumber \\
&& = 0.1193^{+0.0021}_{-0.0019},
\end{eqnarray}
with $\chi^2/$d.o.f.$=43.9/20$ for the C-parameter mean value, where the first (Exp.) and second (Theo.) errors are the experimental and theoretical uncertainties, respectively. Both values are consistent with the world average of $\alpha_s(M^2_Z)=0.1181\pm0.0011$~\cite{Tanabashi:2018oca}.  Since the dominant scale $\mu_r$ uncertainty is eliminated and the convergence of pQCD series is greatly improved after using the PMC, the precision of the extracted $\alpha_s$ values is largely improved. In particular, since a strikingly much faster pQCD convergence is obtained for the thrust mean value~\cite{Wang:2019ljl}, the theoretical uncertainty is even smaller than the experimental uncertainty.

We can also apply the PMC analysis to QED event shapes, where the final-state particles in $e^+ e^- \to \gamma^* \to X(\rm QED)$ are restricted to leptons and photons.  The PMC scales for QCD and QED event shapes are identical at LO after applying the relation between PMC scales: $Q^2_{\rm QCD}/Q^2_{\rm QED}=e^{-5/3}$; this factor converts the scale underlying predictions in the $\overline{\rm MS}$ scheme used in QCD to the scale of the $V$ scheme conventionally used in QED~\cite{Brodsky:1994eh}.  The running of the QED coupling $\alpha(Q^2)$ can be determined from events at a single energy of $\sqrt{s}$~\cite{Abbiendi:2005rx}. Thus one can use the measured event shape distribution in $e^+ e^- \to Z^0 \to X(\rm QED)$ to measure the QED coupling $\alpha(Q^2)$ over a large range of $Q^2$.

\section{Summary}

In summary, the strong running coupling $\alpha_s(Q^2)$ of QCD and its property of asymptotic freedom is fundamental to all QCD analyses; its determination from event-shape distributions is an essential input. The PMC predictions for pQCD are independent of the choice of the  initial renormalization scale and the choice of renormalization scheme. Renormalon divergences are eliminated.  The PMC procedure is identical in the $N_C \to 0$ Abelian limit to the standard Gell-Mann-Low method for QED. It is thus also essential for renormalization scale-setting for grand-unified theories.  We have shown that a comprehensive and self-consistent analysis for both the differential distributions and the mean values for event shapes is obtained by using PMC scale setting.  The highly consistent results for the $T$ and $C$ event-shape distributions verify the applicability of the PMC to pQCD.
The PMC provides a rigorous method for unambiguously setting the renormalization scale as function of the  event-shape kinematics, reflecting the virtuality of the underlying QCD subprocesses.   Thus the PMC provides a remarkable way to verify the running of $\alpha_s(Q^2)$ from the event shape differential measurement at a single energy of $\sqrt{s}$. These new  results for  $\alpha_s(M^2_Z)$ are consistent with the world average and are more precise than the values conventionally obtained from the analysis of event shapes currently used in the world average.

{\bf Acknowledgements}:

This work was supported in part by the Natural Science Foundation of China under Grants No.11625520, No.11705033, No.11847301 and No.11905056; by the Project of Guizhou Provincial Department under Grants No.KY[2016]028 and No.KY[2017]067; and by the Department of Energy Contract No. DE-AC02-76SF00515. SLAC-PUB-17458.


\begin{thebibliography}{99}

\bibitem{Tanabashi:2018oca}
  M.~Tanabashi {\it et al.} [Particle Data Group],
  Phys.\ Rev.\ D {\bf 98}, 030001 (2018).

\bibitem{Dissertori:2009ik}
  G.~Dissertori, A.~Gehrmann-De Ridder, T.~Gehrmann, E.~W.~N.~Glover, G.~Heinrich, G.~Luisoni and H.~Stenzel,
  JHEP {\bf 0908}, 036 (2009).

\bibitem{Abbate:2010xh}
  R.~Abbate, M.~Fickinger, A.~H.~Hoang, V.~Mateu and I.~W.~Stewart,
  Phys.\ Rev.\ D {\bf 83}, 074021 (2011).

\bibitem{Hoang:2015hka}
  A.~H.~Hoang, D.~W.~Kolodrubetz, V.~Mateu and I.~W.~Stewart,
  Phys.\ Rev.\ D {\bf 91}, 094018 (2015).

\bibitem{Ellis:1995jv}
  J.~R.~Ellis, E.~Gardi, M.~Karliner and M.~A.~Samuel,
  Phys.\ Lett.\ B {\bf 366}, 268 (1996).

\bibitem{GellMann:1954fq}
  M.~Gell-Mann and F.~E.~Low,
  Phys.\ Rev.\  {\bf 95}, 1300 (1954).

\bibitem{Brodsky:1997jk}
  S.~J.~Brodsky and P.~Huet,
  Phys.\ Lett.\ B {\bf 417}, 145 (1998).

\bibitem{Brodsky:2011ta}
  S.~J.~Brodsky and X.~G.~Wu,
  Phys.\ Rev.\ D {\bf 85}, 034038 (2012).

\bibitem{Brodsky:2012rj}
  S.~J.~Brodsky and X.~G.~Wu,
  Phys.\ Rev.\ Lett.\  {\bf 109}, 042002 (2012).

\bibitem{Brodsky:2011ig}
  S.~J.~Brodsky and L.~Di Giustino,
  Phys.\ Rev.\ D {\bf 86}, 085026 (2012).

\bibitem{Mojaza:2012mf}
  M.~Mojaza, S.~J.~Brodsky and X.~G.~Wu,
  Phys.\ Rev.\ Lett.\  {\bf 110}, 192001 (2013).

\bibitem{Brodsky:2013vpa}
  S.~J.~Brodsky, M.~Mojaza and X.~G.~Wu,
  Phys.\ Rev.\ D {\bf 89}, 014027 (2014).

\bibitem{Brodsky:2012ms}
  S.~J.~Brodsky and X.~G.~Wu,
  Phys.\ Rev.\ D {\bf 86}, 054018 (2012).

\bibitem{Wu:2014iba}
  X.~G.~Wu, Y.~Ma, S.~Q.~Wang, H.~B.~Fu, H.~H.~Ma, S.~J.~Brodsky and M.~Mojaza,
  Rept.\ Prog.\ Phys.\  {\bf 78}, 126201 (2015).

\bibitem{Wu:2019mky}
  X.~G.~Wu, J.~M.~Shen, B.~L.~Du, X.~D.~Huang, S.~Q.~Wang and S.~J.~Brodsky,
  arXiv:1903.12177 [hep-ph].

\bibitem{Brodsky:1982gc}
  S.~J.~Brodsky, G.~P.~Lepage and P.~B.~Mackenzie,
  Phys.\ Rev.\ D {\bf 28}, 228 (1983).

\bibitem{Brandt:1964sa}
  S.~Brandt, C.~Peyrou, R.~Sosnowski and A.~Wroblewski,
  Phys.\ Lett.\  {\bf 12}, 57 (1964).

\bibitem{Farhi:1977sg}
  E.~Farhi,
  Phys.\ Rev.\ Lett.\  {\bf 39}, 1587 (1977).

\bibitem{Parisi:1978eg}
  G.~Parisi,
  Phys.\ Lett.\ B {\bf 74}, 65 (1978).

\bibitem{Donoghue:1979vi}
  J.~F.~Donoghue, F.~E.~Low and S.~Y.~Pi,
  Phys.\ Rev.\ D {\bf 20}, 2759 (1979).

\bibitem{Catani:1996jh}
  S.~Catani and M.~H.~Seymour,
  Phys.\ Lett.\ B {\bf 378}, 287 (1996).

\bibitem{Ridder:2014wza}
  A.~Gehrmann-De Ridder, T.~Gehrmann, E.~W.~N.~Glover and G.~Heinrich,
  Comput.\ Phys.\ Commun.\  {\bf 185}, 3331 (2014).

\bibitem{Weinzierl:2009ms}
  S.~Weinzierl,
  JHEP {\bf 0906}, 041 (2009).

\bibitem{Chetyrkin:2000yt}
  K.~G.~Chetyrkin, J.~H.~Kuhn and M.~Steinhauser,
  Comput.\ Phys.\ Commun.\  {\bf 133}, 43 (2000).

\bibitem{Wang:2019ljl}
  S.~Q.~Wang, S.~J.~Brodsky, X.~G.~Wu and L.~Di Giustino,
  Phys.\ Rev.\ D {\bf 99}, 114020 (2019).

\bibitem{Heister:2003aj}
  A.~Heister {\it et al.} [ALEPH Collaboration],
  Eur.\ Phys.\ J.\ C {\bf 35}, 457 (2004).

\bibitem{GehrmannDeRidder:2007hr}
  A.~Gehrmann-De Ridder, T.~Gehrmann, E.~W.~N.~Glover and G.~Heinrich,
  JHEP {\bf 0712}, 094 (2007).

\bibitem{Hofman:2008ar}
  D.~M.~Hofman and J.~Maldacena,
  JHEP {\bf 0805}, 012 (2008).

\bibitem{Kramer:1990zt}
  G.~Kramer and B.~Lampe,
  Z.\ Phys.\ A {\bf 339}, 189 (1991).

\bibitem{Gehrmann:2014uva}
  T.~Gehrmann, N.~H\"{a}fliger and P.~F.~Monni,
  Eur.\ Phys.\ J.\ C {\bf 74}, 2896 (2014).

\bibitem{Hoang:2014wka}
  A.~H.~Hoang, D.~W.~Kolodrubetz, V.~Mateu and I.~W.~Stewart,
  Phys.\ Rev.\ D {\bf 91}, 094017 (2015).

\bibitem{Becher:2008cf}
  T.~Becher and M.~D.~Schwartz,
  JHEP {\bf 0807}, 034 (2008).

\bibitem{GehrmannDeRidder:2009dp}
  A.~Gehrmann-De Ridder, T.~Gehrmann, E.~W.~N.~Glover and G.~Heinrich,
  JHEP {\bf 0905}, 106 (2009).

\bibitem{Weinzierl:2009yz}
  S.~Weinzierl,
  Phys.\ Rev.\ D {\bf 80}, 094018 (2009).

\bibitem{Pahl:2007zz}
  C.~J.~Pahl,
  CERN-THESIS-2007-188;
  http://cds.cern.ch/record/2284229

\bibitem{Abbiendi:2004qz}
  G.~Abbiendi {\it et al.} [OPAL Collaboration],
  Eur.\ Phys.\ J.\ C {\bf 40}, 287 (2005).

\bibitem{Pahl:2008uc}
  C.~Pahl {\it et al.} [JADE Collaboration],
  Eur.\ Phys.\ J.\ C {\bf 60}, 181 (2009);
  Erratum: [Eur.\ Phys.\ J.\ C {\bf 62}, 451 (2009)].

  \bibitem{Brodsky:1994eh}
  S.~J.~Brodsky and H.~J.~Lu,
  Phys.\ Rev.\ D {\bf 51}, 3652 (1995).

\bibitem{Abbiendi:2005rx}
  G.~Abbiendi {\it et al.} [OPAL Collaboration],
  Eur.\ Phys.\ J.\ C {\bf 45}, 1 (2006).

\end{thebibliography}
\end{document}